\documentclass[11pt, reqno]{amsart}
\usepackage{amsmath,amsthm,amssymb}

\theoremstyle{plain}
\newtheorem{thm}{Theorem}[section]
\newtheorem{prop}[thm]{Proposition}
\newtheorem{lem}[thm]{Lemma}
\newtheorem{cor}[thm]{Corollary}

\theoremstyle{definition}

\newtheorem{rem}[thm]{Remark}
\newtheorem{defn}[thm]{Definition}
\newtheorem{eg}[thm]{Example}
\newtheorem{subtitle}[thm]{}
\newtheorem{ex}{Exercise}[section]
\numberwithin{equation}{section}

\def\a{\alpha}
\def\b{\beta}

\def\g{\gamma}

\def\l{\lambda}
\def\L{\Lambda}
\def\n{\,\vert\,}

\def\cg{{\mathcal{G}}}

\def\cn{{\mathcal{N}}}

\def\n{\ \vert\ }
\def\tr{{\rm tr}}
\def\bs{\bigskip}
\def\ms{\medskip}
\def\ss{\smallskip}

\def\ni{\noindent}
\def\ti{\tilde}
\def\p{\partial}

\def\I{{\rm I\/}}

\def\diag{{\rm diag}}
\def\ad{{\rm ad}}
\def\Ad{{\rm Ad}}

\def\rd{{\rm d\,}}

\def\R{\mathbb{R} }
\def\C{\mathbb{C}}

\newcommand{\beg}{\begin{eg}}
\newcommand{\eeg}{\end{eg}}
\newcommand{\bthm}{\begin{thm}}
\newcommand{\ethm}{\end{thm}}
\newcommand{\bprop}{\begin{prop}}
\newcommand{\eprop}{\end{prop}}
\newcommand{\bcor}{\begin{cor}}
\newcommand{\ecor}{\end{cor}}
\newcommand{\blem}{\begin{lem}}
\newcommand{\elem}{\end{lem}}
\newcommand{\bca}{\begin{cases}}
\newcommand{\eca}{\end{cases}}
\newcommand{\brem}{\begin{rem}}
\newcommand{\erem}{\end{rem}}
\newcommand{\bpm}{\begin{pmatrix}}
\newcommand{\epm}{\end{pmatrix}}
\newcommand{\bbm}{\begin{bmatrix}}
\newcommand{\ebm}{\end{bmatrix}}
\newcommand{\bvm}{\begin{vmatrix}}
\newcommand{\evm}{\end{vmatrix}}
\newcommand{\bdefn}{\begin{defn}}
\newcommand{\edefn}{\end{defn}}
\newcommand{\bsub}{\begin{subtitle}}
\newcommand{\esub}{\end{subtitle}}
\newcommand{\bex}{\begin{ex}}
\newcommand{\eex}{\end{ex}}
\newcommand{\ben}{\begin{enumerate}}
\newcommand{\een}{\end{enumerate}}

\newcommand{\balign}{\begin{align}}
\newcommand{\ealign}{\end{align}}
\newcommand{\baligns}{\begin{align*}}
\newcommand{\ealigns}{\end{align*}}
\newcommand{\beq}{\begin{equation}}
\newcommand{\eeq}{\end{equation}}

\def\calA{{\mathcal A}}
\def\calC{{\mathcal C}}
\def\calL{{\mathcal L}}

\def\calB{{\mathcal B}}
\def\calM{{\mathcal M}}
\def\calN{{\mathcal N}}
\def\calG{{\mathcal G}}
\def\calJ{{\mathcal J}}
\def\calO{{\mathcal O}}

\def\Tr{{\rm Tr\/}}

\begin{document}

\title[The $n\times n$ KdV hierarchy]
{The $n\times n$ KdV hierarchy}
\author{Chuu-Lian Terng$^\dag$}\thanks{$^\dag$Research supported
in  part by NSF Grant DMS-0707132}
\address{Department of Mathematics\\
University of California at Irvine, Irvine, CA 92697-3875.  Email: cterng@math.uci.edu}
\author{Karen Uhlenbeck$^*$}\thanks{$^*$Research supported in part by the Sid Richardson
Regents' Chair Funds, University of Texas system
\/}
\address{The University of Texas at Austin\\ Department of Mathematics, RLM 8.100\\ Austin, TX 78712. Email:uhlen@math.utexas.edu}



\dedicatory{Dedicated to Dick Palais on the occassion of  his 80th Birthday}

\begin{abstract} 

We introduce two new soliton hierarchies that are generalizations of the KdV hierarchy.  Our hierarchies are restrictions of the AKNS $n\times n$ hierarchy coming from two unusual splittings of the loop algebra.  These splittings come from automorphisms of the loop algebra instead of automorphisms of $sl(n,\C)$. The second flow in the hierarchy is a system of coupled non-linear Schr\"odinger equations. Since they are constructed from a Lie algebra splitting, the general method gives formal inverse scattering, bi-Hamiltonian structures, commuting flows, and B\"acklund transformations for these hierarchies.  
\end{abstract}

\maketitle

\centerline{\bf Introduction}
\ms

While an exact definition of an integrable system is illusive, mathematicians
generally agree that the hallmarks of the the theory are an infinite
number of commuting flows, formal direct and inverse scattering theory,
symplectic structures and B\"acklund transformations.
 More recently, because of the connections with quantum
cohomology, one might add tau functions and actions of Virasoro algebra.  From our
viewpoint we take the basic model formulated  by Ablowitz, Kaul,
Newell and Segur (AKNS), the $n\times n$ model generalizing the $2\times 2$ non-linear
Schr\"oedinger hierarchy, as the basic construction.

There are a number of possible restrictions of this theory in which the
general hierarchies described by AKNS restrict to more specialized
hierarchies. These include the $n$-wave hierarchy, the Kuperschmidt-Wilson hierarchy, and the modified
KdV hierarchy defined by Drinfel'd and Sokolov.  There are several
models in which $n\times n$ matrices are replaced by simple Lie Algebras,
and the formalism carries over. However, all these hierarchies can be obtained from a splitting of loop algebras so that formal scattering, bi-Hamiltonian, commuting conservation laws, B\"acklund transformations, and tau functions can be constructed in a unified way.  In other words, if we find a splitting that has the given equation as one of the flow equations in the hierarchy constructed from the splitting, then we have found the hidden symmetry of this equation. One goal of this paper is to find the symmetry which gives the KdV and Gelfand-Dikki (GD) hierarchies.

In this paper we find two splittings of the Lie algebra of power
series with values in $sl(n,\C)$, which are defined in a different style from the previous examples, to generate two apparently
new hierarchies, which we call the {\it $n\times n$ KdV hierarchy\/} and the {\it $2n\times 2n$ KdV-II hierarchy\/} respectively. The Kuperschmidt-Wilson and the $n\times n$ mKdV flows constructed by Drinfel'd and Sokolov turn out to be equivalent. In fact, they  come from equivalent splittings of Lie algebras.  The $n\times n$ KdV hierarchy is also equivalent to the GD-hierarchy. In fact,  we find a bijection between the phase spaces of these two hierarchies  such that the flows correspond under the bijection. The proof of this equivalence is quite complicated.  In two later papers \cite{TerUhl09a} and \cite{TerUhl09b}, we will give a proof of this equivalence, use it to show that the natural Virasoro symmetry arising from the formal scattering data of the $n\times n$ KdV flows corresponds to the Virasoro
symmetry of GD flows, and use Wilson's tau function for $\calG$ANKS hierarchy \cite{Wil91} to construct tau functions for the $n\times n$ KdV hierarchy and show that solutions can be computed from the tau functions. 

 The plan of the
paper is as follows:  Section 1  reviews a general construction of a hierarchy of commuting flows from a Lie algebra splitting, the Local Factorization Theorem, and formal inverse scattering.  We also
describe the basic examples of the theory, which include the AKNS hierarchy, the Kuperschmidt-Wilson hierarchy, and the $n\times n$ mKdV hierarchy.  In
section 2 we give a splitting of the loop algebras in $sl(n,C)$ that gives the $n\times n$ KdV hierarchy. In section 3 we construct a second splitting to generate the $2n\times 2n$ KdV-II hierarchy.  

\bs

\section{A general method of constructing soliton hierarchy}\label{ec}

We review the method that generates a hierarchy of commuting flows from a splitting of a Lie algebra.  

\bdefn
Let $L$ be a formal Lie group, $\calL$ its Lie algebra, and $L_\pm$ subgroups of $L$ with Lie subalgebra $\calL_\pm$.  The pair $(\calL_+, \calL_-)$ is called a {\it splitting\/} of $\calL$ if
$\calL= \calL_+ + \calL_-$ as direct sum of linear subspaces and $L_+\cap L_-=\{e\}$, where $e$ is the identity in $L$. We call $\calC=(L_+L_-)\cap (L_-L_+)$ the {\it big cell\/} of $L$. In other words, $f$ is in the big cell if and only if $f$ can be factored uniquely as $f_+f_-$ and $g_-g_+$ with $f_\pm, g_\pm\in L_\pm$.  
\edefn

\bthm\label{bv} {\bf (Local Factorization Theorem)}\par

Suppose $L$ is a closed subgroup of the group of Sobolev $H^k$-loops in a finite dimensional Lie group $G$ with $k>\frac{1}{2}$, $L_\pm$ subgroups of $L$ such that $(\calL_+, \calL_-)$ is a splitting of the Lie algebra $\calL$, where $\calL$ and $ \calL_\pm$ are Lie subalgebras of $L$ and $L_\pm$ respectively.  Let $\calO$ be an open subset in $\R^N$ containing the origin, and $g:\calO\to L$ a map such that $(t, \l)\mapsto g(t)(\l)$ is smooth.  Suppose there exists $p_0\in \calO$ such that $g(p_0)= k_+k_-= h_-h_+$ with $k_\pm, h_\pm\in L_\pm$. Then there exist an open subset $\calO_0\subset \calO$ containing $p_0$ and unique $f_\pm, g_\pm : \calO_0\to L_\pm$ such that $g= g_+g_-= f_-f_+$ on $\calO_0$ and $g_\pm(p_0)= k_\pm$, $f_\pm(p_0)= h_\pm$.
\ethm

\begin{proof} 
First we write down the first order systems for $g_+$ and $g_-$.
If $g=g_+g_-$ with $g_\pm\in L_\pm$, then 
$$\a= g^{-1}\rd g= g_-^{-1}\rd g_- + g_-^{-1} g_+^{-1}\rd g_+ g_-,$$ which implies that
$g_-\a g_-^{-1}=\rd g_- g_-^{-1} + g_+^{-1} \rd g_+$.  
Hence 
$$(\rd g_-) g_-^{-1}= \pi_-(g_-\a g_-^{-1}),\quad
g_+^{-1} \rd g_+ = \pi_+( g_-\a g_-^{-1}),$$
where $\pi_\pm$ is the projection of $\calL$ onto $\calL_\pm$ with respect to $\calL=\calL_++\calL_-$. 
Or equivalently,  
\beq\label{bw}
\bca\rd g_-= \pi_-(g_-\a g_-^{-1}) g_-,\\ \rd g_+ = g_+ \pi_+(g_-\a g_-^{-1}).
\eca
\eeq
Note $\a= g^{-1}\rd g$ is given and \eqref{bw} is a first order PDE for $g_-, g_+$ with values in $L_-$ and $L_+$ respectively. Let $L(G)$ denote the group of Sobolev $H^k$ maps from $S^1$ to $G$ with $k>\frac{1}{2}$. Then $L(G)$ is a Hilbert Lie group (cf. \cite{Pal63}) and hence any closed subgroup of $L(G)$ is also a Hilbert Lie group. The Frobenius Theorem for first order system holds for Hilbert manifold.  

Set 
$$\xi= g_-\a g_-^{-1}, \quad \xi_\pm= \pi_\pm(\xi).$$ The integrability condition for system \eqref{bw} is 
$$\rd \xi_-= \xi_-\wedge \xi_-, \qquad \rd \xi_+ =-\xi_+\wedge \xi_+.$$
But a direct computation gives
\begin{align*}
\rd \xi_-& =(\rd \xi)_-=(\rd (g_-\a g_-^{-1}))_-\\
&= (\rd g_- \wedge\a g_-^{-1}+ g_-\a \wedge g_-^{-1}\rd g_- g_-^{-1}+ g_-\rd \a g_-^{-1})_-\\
&= (\xi_- \wedge \xi + \xi \wedge\xi_- + g_-\rd \a g_-^{-1})_-\\
&= (\xi_-\wedge\xi + \xi\wedge\xi_- - g_-\a\wedge\a g_-^{-1})_-\\
& = (\xi_-\wedge\xi + \xi\wedge\xi_- - \xi\wedge \xi)_- = \xi_-\wedge \xi_-.
\end{align*}
A similar computation gives $\rd \xi_+=-\xi_+\wedge\xi_+$.   Hence by the Frobenius Theorem there exist an open neighborhood $\calO_0$ of $p_0$ and $g_\pm:\calO_0\to L_\pm$ such that $g_\pm(p_0)= k_\pm$ and $g_\pm$ satisfy \eqref{bw}.
A direct computation shows that if $g_+, g_-$ solves \eqref{bw} then  
$$(g_+g_-)^{-1}\rd (g_+g_-)= \a= g^{-1} \rd g.$$  
Since $g_+g_-$ and $g$ satisfies the same first order system and agree at $p_0$, $g= g_+g_-$.  Similarly, there exist $f_\pm\in L_\pm$ such that $f_\pm(p_0)= h_\pm$ and $g= f_-f_+$. 
\end{proof}

\ss

\ni{\bf Birkhoff Factorization} \hfil

Let $L$ be the group of $H^k$ maps from $S^1$ to $GL(n,\C)$, $L_+$ the subgroup of $f\in L$ such that $f$ is the boundary value of a holomorphic map defined on $|\l |<1$, and $L_-$ the subgroup of $f\in L$ such that $f$ extends holomorphically to $\infty\geq |\l | >1$ and is equal to $\I$ at $\l=\infty$.  Let $\calL, \calL_\pm$ denote the corresponding Lie algebras.  Then $(\calL_+, \calL_-)$ is a splitting and Theorem \ref{bv} is a consequence of the Birkhoff Factorization Theorem (cf. \cite{PreSeg86}).  Moreover, the big cell is open and dense in $L$.  

\ss

\bdefn  A  {\it vacuum sequence\/} of a splitting $(\calL_+, \calL_-)$ of $\calL$ is a linearly independent, commuting sequence $\calJ=\{J_j\n j\, {\rm an\, integer}, j\geq 1\}$ in $\calL_+$ such that $J_j= \phi_j(J_1)$ for some analytic function $\phi_j$ in the enveloping algebra of $\calL$ for all $j\geq 1$.
\edefn   

\ms
\ni {\bf Commuting flows on $L_-$}.

Let $\pi_\pm$ denote the projection of $\calL$ onto $\calL_\pm$ with respect to the sum $\calL=\calL_++ \calL_-$.  We call the following flow equation on $L_-$,
\beq\label{bt}
\frac{\p M}{\p t_j}M^{-1}= -\pi_-(MJ_jM^{-1}),
\eeq
{\it the flow on $L_-$ generated by $J_j$}.
Use $[J_j, J_k]=0$ for all $i,j$ and a direct computation to get
$$-(P_j)_{t_k} + (P_k)_{t_j}= [P_j, P_k], \qquad {\rm where\/} \quad P_j= -\pi_-(MJ_jM^{-1}).$$
So flows \eqref{bt} defined on $L_-$ commute.

\ms
\ni {\bf Soliton hierarchy}\hfil\par

The above flows \eqref{bt} on $L_-$ give rise to explicit partial differential equations on 
\beq\label{by}
\calM=\{\pi_+(gJ_1g^{-1})\n g\in L_-\}.
\eeq
 The phase space of the soliton flows constructed from the splitting $(\calL_+, \calL_-)$ and the vacuum sequence $\{J_j\n j\geq 1\}$ is $C^\infty(\R, \calM)$.  

Assume that given a smooth $\xi:\R\to \calM$, there is a unique $Q_j(\xi)\in \calL$ satisfying the following conditions:
\beq\label{ag}
\bca [\p_x-\xi, Q_j(\xi)]=0,\\ 
Q_j(\xi) \, {\rm is\,  conjugate\,  to\,}\,  J_j \,\, {\rm by\,\, } L_-,\\
Q_j(J_1)= J_j.\eca
\eeq
Note that 
$$ [\p_x- \xi, (Q_j(\xi))_+]= -[\p_x-\xi, (Q_k(\xi))_-]_+ = [\xi,  (Q_k(\xi))_-]_+,$$
which is tangent to $C^\infty(\R, \calM)$.  Hence 
\beq\label{bx}
\frac{\p \xi}{\p t_j} = [\p_x- \xi, (Q_j(\xi))_+].
\eeq
defines a flow on $C^\infty(\R, \calM)$.  We call \eqref{bx} the {\it flow generated by $J_j$\/}.
The collection of these flows is called the {\it soliton hierarchy constructed from $(\calL_+, \calL_-)$ and $\{J_j\n j\geq 1\}$}.
If $Q_j(\xi)$ depends only on $\xi$ and its derivatives, then \eqref{bx} is a PDE.

\ss
Note that $J_1\in \calM$.
The tangent space of $\calM$ at $J_1$ is 
$$Y:=\pi_+([J_1,\calL_-]).$$   In examples given below, $\calL$ is a subalgebra of the algebra of loops in $gl(n)$, $J_1$ is of the form $a\l +b$ for some constants $a, b\in gl(n)$,  $Y$ is a finite dimension linear subspace of $gl(n)$ (constant loops), and  $\calM$ is the affine space 
$$\calM= J_1+ Y.$$ 
\ss

 Next we outline a method of computing these flows when $\calL$ is a subalgebra of the Lie algebra of loops in $gl(n)$.
Expand $Q_j(\xi)$ in $\l$:
 $$Q_j(\xi)(\l)= \sum_k Q_{j,k} \l^k.$$
Equating the coefficients of $\l^j$ in  $[\p_x-\xi, Q_j(\xi)]=0$ give recursive formulas for $Q_{j,k}$'s. The second condition of \eqref{ag} holds if and only if $Q_j$ and $J_j$ have the same minimal polynomials. These give another set of equations for $Q_{j,k}$'s.  It often is the case that these two sets of formulas determine $Q_{j,k}$'s.   Since $J_j=\phi_j(J_1)$, $Q_j= \phi_j(Q_1)$, we only need to compute $Q_1$.
 
\ms 
The proof of the next Proposition is straight-forward.

\bprop\label{ac} 
The following statements are equivalent for $\xi: \R^2\to \calM$:
\ben 
\item[(1)] $\xi$ is a solution of the flow \eqref{bx} generated by $J_j$,
\item[(2)] $[\p_x - \xi, \, \p_{t_j} - (Q_j(\xi))_+]=0$,
\item[(3)] the following linear system is solvable for any initial data $c\in L_+$:
\beq\label{ae} E_xE^{-1}= \xi, \quad E_{t_j} E^{-1} = (Q_j(\xi))_+, \quad E(0,0)= c.\eeq
\een
\eprop  

\ni{\bf Frames}\par

We call the solution of \eqref{ae} with initial data $c=\I$ the {\it frame\/} of $\xi$.  The constant map $\xi= J_1$ is a solution of the flow generated by $J_j$ for all $j\geq 1$, and is called the {\it vacuum solution\/} and its frame  
$$V(x,t_j)= \exp(xJ_1+ t_j J_j)$$ is called the {\it vacuum frame\/}.  

\ss
\ni {\bf Baker function}

Given $\xi\in C^\infty(\R, \calM)$, assume that there is a $M(x)\in L_-$ such that  
$$M(\p_x- J_1)M^{-1}= \p_x -\xi.$$
Or equivalently, 
\begin{equation}\label{bs}
M_xM^{-1} + MJ_1M^{-1}= \xi.
\end{equation}
Since $M_xM^{-1}\in L_-$, we have
\beq\label{bj}
\xi= \pi_+(MJ_1M^{-1}), \qquad M_xM^{-1}= -\pi_-(MJ_1M^{-1}).
\eeq 
 Such $M$ is usually called a {\it Baker function\/} or the {\it bare wave function\/} of $\p_x- \xi$, and $f=M(0)\in L_-$ a {\it formal scattering data\/} for the operator $\p_x-\xi$.  
 
 Let $A$ denote the centralizer of $J_1$ in $L_-$.  If $M$ is a Baker function of $\p_x-\xi$ and $g\in A$ a constant, then $Mg$ is a again a Baker function of $\p_x-\xi$.  Hence we call the right coset $[M(0)]$ in $L_-/A$ the {\it scattering coset} of $\p_x-\xi$ in \cite{TerUhl98}.  
 
 If $\xi= J_1+u$ and $u$ has compact support, then it follows from results in \cite{BeaCoi84} and {BDT88} that there is a Baker function $M$. The results in these papers on rapidly decaying $u$ apply to our examples as well. 
 
 \ss
 
 Suppose $\p_x-\xi$ has a Baker function $M$. Set $\eta_j= MJ_jM^{-1}$. Since $[\eta_1, \eta_j]=0$ and $\xi= (\eta_1)_+$, we have 
\begin{align*}
[\p_x-\xi, \eta_j] &= [M_xM^{-1}, \eta_j] -[\xi, \eta_j]= -[(\eta_1)_-, \eta_j] -[(\eta_1)_+, \eta_j]\\
& = -[\eta_1, \eta_j]=0.
\end{align*}
It follows that $MJ_jM^{-1}$ satisfies \eqref{ag}.  So the flow generated by $J_j$ can be written as 
\beq\label{aj}
((MJ_1M^{-1})_+)_{t_j} =[\p_x -(MJ_1M^{-1})_+, (MJ_jM^{-1})_+].
\eeq
Here we use the notation $\eta_+=\pi_+(\eta)$. 
Note that if $M'$ is another Baker function for $\p_x-\xi$, then $M'J_j (M')^{-1}= MJ_j M^{-1}$. So the flow equation \eqref{aj} is independent of the choice of Baker functions. 
 
 \bprop\label{ai}
 Suppose $M(x,t)$ is a solution of the first and the $j$-th flows \eqref{bt} on $L_-$. Then $\xi=(MJ_1M^{-1})_+$ is a solution of the flow \eqref{bx} generated by $J_j$.
 \eprop
 
 \begin{proof}
 It follows from the definition of $\xi$ that $M$ is a Baker function of $\xi$.  Set $\eta_i=MJ_iM^{-1}$ for $i=1, j$. Then
 $$\frac{\p\xi}{\p t}= [M_tM^{-1}, (MJ_1M^{-1})]_+= -[(\eta_j)_-,\eta_1]_+.$$
 Since
 \begin{align*}
 &[\p_x-\xi,  (\eta_j)_+] = -[(\eta_1)_-, \eta_j]_+ -[\xi, (\eta_j)_+] \\
 &= -[(\eta_1)_-, (\eta_j)_+]_+ -[\xi, (\eta_j)_+] = -[\eta_1, (\eta_j)_+]_+ = -[(\eta_j)_-, \eta_1]_+,
 \end{align*}
 then this is equal to $\frac{\p\xi}{\p t}$.  
 \end{proof}

\ms
\ni {\bf Formal inverse scattering\/} \cite{TerUhl98}\hfil\par
 
Given $f\in L_-$, by Theorem \ref{bv} we can factor 
$$V(x,t_j)f^{-1}= M^{-1}(x,t_j)E(x,t_j)$$ with $M(x,t_j)\in L_-$ and $E(x,t_j)\in L_+$, where $V(x,t_j)=\exp(xJ_1+ t_jJ_j)$ is the vacuum frame of the $j$-th flow.    Since $E= MVf^{-1}$, we have
\begin{align*}
&E_xE^{-1}= M_xM^{-1}+ MV_xV^{-1}M^{-1}= M_x M^{-1}+ MJ_1 M^{-1},\\
&E_{t_j}E^{-1}= M_{t_j}M^{-1}+ MV_{t_j} V^{-1} M^{-1}= M_{t_j}M^{-1}+ MJ_j M^{-1}.
\end{align*}
 But $E_{t_j}E^{-1}\in \calL_+$ and $M_{t_j}M^{-1}\in \calL_-$ imply that 
 \begin{subequations}
 \begin{gather}
  E_xE^{-1}= \pi_+(MJ_1M^{-1}), \quad E_{t_j}E^{-1}= \pi_+(MJ_jM^{-1}), \label{am}\\
  M_xM^{-1}= -\pi_-(MJ_1M^{-1}), \quad M_{t_j} M^{-1}= -\pi_-(MJ_j M^{-1}).\label{ba}
 \end{gather}
 \end{subequations}
 It follows from \eqref{ba} that $M$ solves the first and the $j$-th flow equation \eqref{bt} on $L_-$. By Proposition \ref{ai}, $\xi_f:=\pi_+(MJ_1M^{-1})$ is a solution of the $j$-th flow \eqref{aj},  $M$ is the Baker function of $\p_x-\xi$, and $f=M(0)$. 
  Equation \eqref{am} implies that $E$ is the frame for $\xi_f$.

We call the above construction of solution $\xi_f$ from $f$  the {\it formal inverse scattering\/}.  This formal inverse scattering solves the Cauchy problem for all flows as follows: Given an initial data $\xi_0:\R\to \calM$, assume that there is a bare wave function $M_0$ for $\p_x-\xi_0$. The {\it scattering data\/} of $\xi_0$ is $f=M_0(0)$. Then $\xi_f$ is a solution of the $j$-th flow with $\xi_f(x,0)= \xi_0(x)$.

\ms
\ni{\bf Action of $L_-$} (cf. \cite{TerUhl98}, \cite{TerUhl00a})\hfil

Let $\xi$ be a solution of the flow  \eqref{bx} generated by $J_j$ defined in an open subset $\calO$ of $\R^2$ containing the origin, and suppose $E$ is its frame.  So $E(x,t)\in L_+$.   
Given $f\in L_-$, by the Local Factorization Theorem \ref{bv}, there exists an open subset $\calO_0\subset \calO$ containing the origin such that $fE(x,t) = \ti E(x, t)\ti f(x,t)$ with $\ti E\in L_+$ and $\ti f\in L_-$ for all $(x,t)\in \calO_0$. Then
\ben
\item  $\ti \xi$ is a new solution of the flow \eqref{bx} generated by $J_j$, and $\ti E$ is the frame of $\ti\xi$,
\item $f\ast\xi= \ti \xi$ defines an action of $L_-$ on the space of germs of solutions of the flow equation \eqref{bx} generated by $J_j$ at the origin, 
\item if $f\in L_-$ is a rational map, then $f\ast \xi$ can be obtained algebraically from $E$ via residue calculus,
\item if $f\in L_-$ is rational and has minimal number of poles, then $\xi\mapsto f\ast\xi$ often gives B\"acklund or Darboux transforms,
\item the orbit of the vacuum solution $J_1$ under the group of rational maps in $L_-$ can be computed explicitly, and they are the soliton solutions. 
\een

Example \ref{ab} - \ref{dt} are well-known soliton hierarchies which can be constructed using this splitting method. There are many references to these in the literature.

\ss
\beg \label{ab} {\bf The $n\times n$ AKNS flows (or the $SL(n,\C)$-hierarchy)\/} \cite{AKNS74}\hfil\par

Let $a_1, \ldots, a_{n-1}$ be a basis of the space of diagonal matrices in $sl(n,\C)$ such that  $a_1$ has distinct eigenvalues.  Let $L(SL(n))$ denote the group of  smooth loops $f:S^1\to SL(n,\C)$, $L_+(SL(n))$ the subgroup of $f\in L(SL(n))$ that can be extended holomorphically to $|\l|<1$, and $L_-(SL(n))$ the subgroup of $f\in L(SL(n))$ that can be extended holomorphically to $\infty=|\l|>1$ and $f(\infty)=\I$.  The corresponding Lie algebras are 
\begin{align*}
\calL(sl(n)) &= \{A(\l)=\sum_i A_i \l^i: S^1\to sl(n,\C) \,\, {\rm smooth\/} \n A_i\in sl(n,\C)\},\\
\calL_+(sl(n))&=\{A\in \calL(sl(n))\n A(\l)=\sum_{j\geq 0} A_j \l^j\},\\
\calL_-(sl(n))&=\{A\in \calL(sl(n))\n A(\l)=\sum_{j<0}A_j \l^j\}.
\end{align*}
It is easy to check that  $(\calL_+(sl(n)), \calL_-(sl(n)))$ is a splitting of $\calL(sl(n))$.  The countable set 
$$\calJ=\{a_i\l^j\n 1\leq i\leq n-1, j\geq 1\}$$
is a vacuum sequence of this splitting with $J_1= a_1\l$. A direct computation implies that $\calM=a_1\l + Y$, where
$$Y=\pi_+([a_1\l, \calL_-])= \{u=(u_{ij})\in sl(n,\C)\n u_{ii}=0, \, \, \forall\,\, 1\leq i\leq n\}.$$ The hierarchy of flows constructed from this splitting and $\calJ$ is the $n\times n$ AKNS hierarchy. 
\eeg

 \beg  {\bf The $SU(n)$-hierarchy\/}  \cite{TerUhl00a}\hfil\par
 
 Let $\calL(sl(n))$ and $\calL_\pm(sl(n))$ be as in the $n\times n$ AKNS flows, $\tau$ the involution of $SL(n,\C)$ defined by $\tau(g)= (\bar g^t)^{-1}$, and $\tau_\ast$ the induced involution on $sl(n,\C)$, i.e.,  $\tau_\ast(\xi)= -\xi^* = -\bar\xi^t$.  Let $\hat \tau_\ast$ be the involution on $\calL(sl(n))$ defined by
 $$\hat\tau_\ast(A)(\l)= -\tau_\ast(A(\bar \l)) = - (A(\bar \l))^*.$$
 Let $\calL$ denote the fixed point set of $\hat \tau$. In other words, $A\in \calL$ if and only if $A(\l)= -(A(\bar \l))^*$. Set $\calL_\pm= \calL\cap \calL_\pm(sl(n))$. Then $(\calL_+, \calL_-)$ is a splitting of $\calL$.  Let  $a_1, \ldots, a_{n-1}$ be linearly independent diagonal matrices in $su(n)$ such that $a_1$ has distinct eigenvalues. Set $J_1=a_1\l$, and $\calJ=\{a_i \l^j\n 1\leq i\leq n-1, j\geq 1\}$.  Then $\calJ$ is a vacuum sequence in $\calL_+$ and $\calM=a_1\l + Y$, where
 $$Y=\pi_+([a_1\l, \calL_-])=\{u=(u_{ij})\in su(n)\n u_{ii}=0 \, \, {\rm for\, all\,\, } 1\leq i\leq n\}.$$ The hierarchy of flows constructed from this splitting and $\calJ$ are flows on $C^\infty(\R, Y)$.  When $n=2$, choose $a_1=\diag(i, -i)$. Then the flow generated by $a_1\l^2$ is the non-linear Sch\"rodinger equation (NLS) for $u=\bpm 0 & q\\ -\bar q & 0\epm$: 
 $$q_t= \frac{1}{2}( q_{xx} + 2 |q|^2 q).$$
   For general $n$, the flow generated by $a_2\l$ is the following equation for $u\in Y$:
 $$(u_{ij})_t= \frac{b_i-b_j}{c_i-c_j} (u_{ij})_x + \sum_{k=1}^n\left( \frac{b_k-b_j}{c_k-c_j} -\frac{b_i-b_k}{c_i-c_k}\right) u_{ik} u_{kj},$$
 where $a_1=\diag(c_1, \ldots, c_n)$ and $a_2= \diag(b_1, \ldots, b_n)$.  When $n=3$, this is the $3$-wave equation given in \cite{ZakMan73}. 
  \eeg
  
  \beg{\bf The $\frac{SU(n)}{SO(n)}$-hierarchy\/} \cite{TerUhl00a}\hfil\par
  
  The symmetric space $\frac{SU(n)}{SO(n)}$ is given by two commuting involutions $\tau, \sigma$ of $SL(n,\C)$, where $\tau(g)=(\bar g^t)^{-1}$ and $\sigma(g)= (g^t)^{-1}$.  Note that $\tau_\ast$ is conjugate linear, $\sigma_\ast$ is complex linear, $SU(n)$ is the subgroup of $SL(n,\C)$ fixed by $\tau$, and $SO(n)$ is the subgroup of $SU(n)$ that is fixed by $\sigma$.  Let $\calL$ denote the subalgebra of $A\in \calL(sl(n,\C))$ satisfying the $\frac{SU(n)}{SO(n)}$-reality condition
  $$A(\l)=\tau_\ast(A(\bar\l))= - A(\l)^*, \quad A(\l)= \sigma_\ast(-\l) = -A(-\l)^t.$$
  Let $\calL_\pm= \calL\cap \calL(sl(n,\C))$.  Then $(\calL_+, \calL_-)$ is a splitting of $\calL$.  Let $a_1, \ldots, a_n$ be a basis of diagonal matrices in $su(n)$ such that $a_1$ has distinct eigenvalues, $J_1= a_1\l$, and 
  $$\calJ=\{a_i \l^j\n 1\leq i\leq n-1, j\geq 1 \, {\rm is\, odd\/}\}.$$
  Then $\calJ$ is a vacuum sequence and the corresponding hierarchy is called the $\frac{SU(n)}{SO(n)}$-hierarchy.  When $n=2$ and $a_1=\diag(i, -i)$, the flow generated by $a_1\l^3$ is the mKdV equation.
  \eeg
  
   \ss
  \beg {\bf The matrix NLS hierarchy} \cite{ForKul83}\hfil\par
  
  Let $\calL=\{A\in \calL(gl(n))\n A(\bar\l)^* + A(\l)=0\}$, and $\calL_\pm = \calL\cap \calL_\pm(gl(n))$.  Then $(\calL_+, \calL_-)$ is a splitting of $\calL$.  Let $a=i\diag(1, \ldots, 1, -1, \ldots, -1)$ with eigenvalues $i, -i$ of multiplicities $k$ and $(n-k)$ respectively.  Set $J_j= a\l^j$.  Then $J_j = f_j(J_1)$ for some analytic function $f_j$.  The phase space is $a\l + Y$, where
  $$Y= [a, \calL_-]_+= \left\{ \bpm 0 & q\\ -q^* &0\epm\n q\quad {\rm is\, a\,} k\times (n-k)\, \, {\rm complex\, matrix\/}\right\}.$$
  The second flow is the matrix NLS: 
  $$q_{t_2}= \frac{i}{2}(q_{xx} + 2qq^*q).$$
  \eeg
  
  \ss
  
  \beg{\bf The $G$-, $U$-, $\frac{U}{K}$- hierarchies} \cite{TerUhl00a} \hfil\par
  
  Note that $SU(n)$ is a real form of the complex simple Lie group $SL(n,\C)$, and $\frac{SU(n)}{SO(n)}$ is a symmetric space.  
  We can replace $SL(n, \C)$ by a complex simple Lie group $G$, $SU(n)$ by a real form $U$ of $G$, and $\frac{SU(n)}{SO(n)}$ by a symmetric space $\frac{U}{K}$ to construct $G$-, $U$-, $\frac{U}{K}$- hierarchies (cf. \cite{TerUhl98, TerUhl00a}). 
\eeg

\ss
\beg\label{ds}{\bf The Kupershmidt-Wilson (KW) hierarchy} \cite{KupWil81}\hfil\par

Let  $C$ be the permutation matrix $e_{21}+ e_{32} + \cdots + e_{n,n-1}+ e_{1n}$, and  $L(sl(n))$ and $L_\pm(sl(n))$ as in the $n\times n$ AKNS flows. Let $\sigma$ be the automorphism of $\calL(sl(n))$ defined by 
$$\sigma(A)(\l)= C^{-1}A(\a^{-1} \l)C, \quad  \a=e^{2\pi i/n}. $$
   Let  $\calL$ be the fixed point set of $\sigma$.  In other words, $A\in \calL$ if and only if 
$$C^{-1}A(\a^{-1}\l)C = A(\l).$$
Let  $\calL_\pm= \calL\cap \calL_\pm(sl(n))$, and 
$$a=\diag(1, \a, \a^2, \ldots, \a^{n-1}), \quad J_1= J= a\l.$$  
Then $(\calL_+, \calL_-)$ is a splitting of $\calL$ and $\calJ=\{J^i\n i\geq 1 \, {\rm integer\/}\}$ is a vacuum sequence in $\calL_+$. The space 
$$Y=\pi_+([a\l, \calL_-])=\{\sum_{i=1}^{n-1} u_i C^i\n u_i\in \C\}$$ is of dimension $n-1$. The flows given by this splitting and vacuum sequence are the KW flows.  When $n=2$, this gives exactly the hierarchy of mKdV
$$q_t= \frac{1}{4} (q_{xxx}- 6q^2 q_x).$$  So the KW hierarchy is a natural $n\times n$ generalization of the standard complex mKdV hierarchy.
\eeg

\ss
\beg\label{dt} {\bf The $n\times n$ mKdV hierarchy} (Drinfel'd-Sokolov \cite{DriSok84})\hfil\par

Let $\calL=\calL(sl(n))$, and 
\begin{align*}
\calL_+&=\{A(z)= \sum_{i>0} A_i z^i + A_0\in \calL\n A_0\in \calB_+\},\\
\calL_-&=\{A(z)=\sum_{i<0} A_i z^i + A_0\in \calL\n A_0\in \calN_-\}.
\end{align*}
Here $\calB_+$ is the subalgebra of upper triangular matrices in $sl(n,\C)$ and $\calN_-$ is the subalgebra of strictly lower triangular matrices in $sl(n,\C)$.  
Let $J= e_{n1}z + b$, where $$b= e_{12} + e_{23} + \cdots + e_{n-1, n}.$$
Then $(\calL_+, \calL_-)$ is a splitting of $\calL(sl(n))$, and $\calJ=\{J^j\n j\geq 1, \, {\rm integer}\}\subset \calL_+$ is a vacuum sequence.  The space $Y$ defined above is
$$Y=\pi_+([J, \calL_-])=\{\diag(u_1, \ldots, u_n)\n u_j\in \C, \sum_{i=1}^n u_j=0\}.$$
There is a bare wave functions for $\p_x -(J+ u)$ with $u:\R\to Y$. To see this, we use $$z=\l^n$$ and construct an equivalent splitting in $\l$ loops for the $n\times n$ mKdV.  Let 
$$D(\l)= \diag(\l^{n-1}, \l^{n-2}, \ldots, 1), \quad p= e_{12} + e_{23} + \ldots + e_{n-1,n} + e_{n1}.$$   First note that $J$ is conjugate to $p\l$,
$$J= e_{n1} \l^n + b= D(\l)^{-1}\, p\l\, D(\l).$$
 We use the isomorphism $\Ad(D(\l))$ and $z=\l^n$ to construct an equivalent splitting as follows: Let
\begin{align*}
&\ti \calL(\l)=\{D(\l)B(\l^n) D(\l)^{-1}\n B\in \calL(sl(n))\},\\
&\ti\calL_\pm(\l)=  \{D(\l)B(\l^n) D(\l)^{-1}\n B\in \calL_\pm\}.
\end{align*}
In other words, $\ti\calL(\l)$ is the subalgebra of $A(\l)=\sum_j A_j \l^j$ that satisfies the following reality condition 
$$D(\l)^{-1}A(\l) D(\l) = D(\a\l)^{-1} A(\a\l) D(\a\l),$$
where $\a=e^{2\pi i/n}$.  Thus $A\in \ti \calL_\pm(\l)$ if and only if $D(\l)^{-1} A(\l) D(\l)$ is a power series in $\l^n$ and the constant term is in $\calB_+$ and in $\calN_-$ respectively.
Then $(\ti \calL_+(\l), \ti \calL_-(\l))$ is a splitting of $\ti \calL(\l)$, $\{\ti J^i\n \ti J= p\l, \, i\geq 1\}$ is a vacuum sequence in $\calL_+$, and 
$$\ti Y= \pi_+([\ti J_1, \ti \calL_-])= \{\sum_{j=1}^n u_j e_{jj}\n \sum_j u_j=0\}.$$ 
 Since $p$ is regular, $\p_x -(\ti J + u)$ has a bare wave function $M$ and the general method of constructing soliton hierarchy works. This gives the $n\times n$ mKdV hierarchy.  For example, for $n=2$ we have $u=\diag(q, -q)$, and
 $$M \ti J M^{-1}=\ti J +\sum_{i\leq 0} Q_i \l^i$$
 has the form 
 $$Q_{2i}= \diag(A_{2i}, -A_{2i}), \quad Q_{2i+1}= \bpm 0& B_{2i+1}\\ C_{2i+1} & 0\epm.$$
These $Q_i's$ can be computed from the recursive formula
 $$\p_x Q_i -[u, Q_i]=[p, Q_{i-1}],$$
 and the third flow is the mKdV $q_t= \frac{1}{4}( q_{xxx} - 6 q^2 q_x)$. 

\eeg

\ss
The following was known to Drinfel'd and Sokolov \cite{DriSok84}, but we give a proof anyway.

\bprop\label{du}  The  KW-hierarchy and Drinfel'd-Sokolov $n\times n$ mKdV hierarchy are equivalent.  
\eprop

\begin{proof} 
 We note that:
 \ben
 \item $e_{n1}\l^n+b$ has eigenvalues $\l, \a\l, \ldots, \a^{n-1}\l$ and can be diagonalized 
$$e_{n1}\l^n+b= U(\l) a\l U(\l)^{-1},$$
where $b=e_{12} + e_{23} + \ldots + e_{n-1,n}$, $\a=e^{2\pi i/n}$, $a=\diag(1,\a, \ldots, \a^{n-1})$, and the $ij$-th entry of $U(\l)$ is $U_{ij}(\l)= \a^{(i-1)(j-1)}\l^{i-1}$.  
\item The $ij$-th entry of $U(\l)^{-1}$ is $\a^{-(i-1)(j-1)}\l^{-(j-1)}$.
\item 
$U(\a\l)= U(\l) C, \quad {\rm where\,\,} C= e_{21}+ e_{32} + \cdots + e_{1n}$.
\een
 Let $\calL^{kw}=\calL^{kw}_++\calL^{kw}_-$ denote the splitting that gives the KW-hierarchy as in Example \ref{ds}. Recall that  $A\in \calL^{kw}$ if and only if 
 $CA(\a\l)C^{-1}= A(\l)$.  By (3), we have
$$ U(\a\l) A(\a\l) U(\a\l)^{-1}= U(\l) CA(\a\l) C^{-1} U(\l)^{-1}.$$ 
So $A\in \calL^{kw}$ if and only if  $U(\a\l)A(\a\l) U(\a\l)^{-1}= U(\l) A(\l) U(\l)^{-1}$, 
i.e., $U(\l) A(\l) U(\l)^{-1}$ is a power series in $\l^n$.

Let $\calL= \calL_+ + \calL_-$ denote the splitting that gives the Drinfel'd-Sokolov $n\times n$ mKdV as in Example \ref{dt}.  Define the map $F:\calL\to \calL^{kw}$ by 
$$F(\xi)(\l)= U(\l)^{-1}\xi(\l^n)U(\l).$$  It follows easily from the explicit formulas for $U(\l)$ and $U(\l)^{-1}$  that $F$ maps $\calL_\pm$ isomorphically to $\calL^{kw}_\pm$, maps $e_{n1}z+ b$ to $a\l$, and maps the vacuum sequence of the $n\times n$ mKdV hierarchy to that of the KW-hierarchy.  This shows that these two hierarchies are isomorphic under $F$. 
\end{proof}

\bs
\section{Construction of $n\times n$ KdV flows}\label{ed}

We have given a list of known examples, and we now discuss a new example which has a different flavor. 

The Lax pair for KdV 
$q_t=\frac{1}{4}(q_{xxx}- 6qq_x)$ is 
$$\left[\p_x -(a\l+ u), \, \p_t -(a\l^3+u\l^2+Q_{-1}\l + Q_{-2})\right]=0,$$
where 
\begin{align*}
& a=\bpm 1&0\\ 0&-1\epm, \quad u= \bpm 0&1\\ q&0\epm, \\
& Q_{-1}=\frac{1}{2}\bpm -q& 0\\ -q_x&q\epm, \quad
Q_{-2}=\frac{1}{4}\bpm q_x& -2q\\ q_{xx} - 2q^2 & -q_x\epm.
\end{align*}
Let $\calL(sl(n))$ and $\calL_\pm (sl(n))$ as in Example \ref{ab}.  The phase space of the AKNS $2\times 2$  hierarchy generated by $\{a\l^j\n j\geq 1\}$ is the space of maps $u=\bpm 0 &r\\ q & 0\epm$, and the third flow in the $2\times 2$ AKNS hierarchy is 
$$\bca q_t= \frac{1}{4} (q_{xxx}- 6qrq_x),\\ r_t= \frac{1}{4}(r_{xxx} - 6 qrr_x).\eca$$
  It is known that (cf. \cite{AKNS74}) the subset of $u$ with $r=1$ is magically invariant under the odd flows in the $2\times 2$ hierarchy, which is the KdV hierarchy.  However, it was not clear why this works.  
In an earlier paper \cite{TerUhl00a}, we noted that KdV hierarchy is the hierarchy associated to a splitting of the subalgebra $\calL_2$ of $\xi \in\calL(sl(2))$  satisfying the following reality condition:
\begin{equation}\label{za}
\phi(\l)\xi(\l)\phi(\l)^{-1}= \phi(-\l) \xi(-\l) \phi(-\l)^{-1}, \quad \phi(\l)= \bpm 1&0\\ \l & 1\epm.
\end{equation}
 A direct computation shows that $\xi(\l)=\sum \xi_j \l^j$ lies in $\calL_2$ if and only if 
$$\xi_{2j}= \bpm A_{2j} & B_{2j}\\ C_{2j} & -A_{2j}\epm, \quad \xi_{2j+1}=\bpm B_{2j} & 0 \\ -2 A_{2j} & -B_{2j}\epm$$
for all $j$.  So 
$$(\calL_2)_+=\{\xi\in \calL_2\n \xi(\l)=\sum_{j\geq 0} \xi_j \l^j\}, \quad (\calL_2)_-=\{\xi\in\calL_2\n \xi(\l)= \sum_{j<0} \xi_j\l^j\}$$
are Lie subalgebras of $\calL_2$ and $(\calL_2, (\calL_2)_+, (\calL_2)_-)$ is a splitting.  

Let $J= a\l +e_{12}$. It can be checked easily that $J^2=\l^2\I_2$ and $J^{2j+1}= \l^{2j} J$ for all $j\geq 0$, where $\I_2$ is the $2\times 2$ identity matrix. Then 
$$\{J^{2i+1}\n i\geq 0\}$$ is a vacuum sequence, which generates the KdV hierarchy.  

We also noted in  \cite{TerUhl00a} that  $\phi$ in the reality condition \eqref{za} is the strictly lower triangular matrix in the Bruhat decomposition of the eigen-matrix of $e_{21}\l^2 + e_{12}$.  This led us to consider a natural generalization of $\calL_2$ to  a Lie subalgebra $\calL$ of $\calL(sl(n))$. Let $\calL_\pm =\calL\cap \calL_\pm(sl(n))$.  We claimed in \cite{TerUhl00a} (without a  proof, in fact we thought the proof was obvious) that $(\calL_+, \calL_-)$ is  a splitting of $\calL$ and use it to generate a hierarchy of flows and compute B\"acklund transformations.  Later we realized that the proof that $(\calL_+,\calL_-)$ is a splitting is quite complicated.  The goal of this section is to give the proof of this fact.  We then use the method described in section \ref{ec} to generate the infinite sequence of flows. 

Our splitting potentially gives a general method to construct KdV type hierarchies from the loop algebra of a complex simple Lie algebra $\calG$. However, this construction is not self evident. 
 
 The vacuum $e_{21}\l^2+ e_{12}$ can be viewed as $\l^2$ times the lowest root plus the sum of simple roots of $sl(2)$.  To generalize KdV, it is natural to consider  $e_{n1}\l^n+ b$. Here  $e_{12}, \ldots, e_{n-1, n}$ form a set of simple roots, $e_{n1}$ is the lowest root, and 
 $b= e_{12}+ e_{23} + \ldots + e_{n-1, n}$
  is the sum of simple roots. We will use the strictly lower triangular matrix factor  of the Bruhat factorization of the eigen-matrix of  $e_{n1}\l^n+b$ to construct  Lie subalgebras of $L(sl(n))$ and splittings. These splittings will give not one but two generalizations of the KdV hierarchy. 
  
  We first set up some standard notations.  Let $\cg_k$ denote the set of all $\xi$ in $sl(n,\C)$ with weight $k$, i.e., 
$$\cg_k=\{X=(x_{ij})\in sl(n,\C)\n x_{ij}=0 \ {\rm unless\ } j-i=k\}.$$
Then 
$$[\cg_i, \cg_j]\subset \cg_{i+j}, \quad \cg_i \cg_j\subset \cg_{i+j},$$
and $\cg_i=0$ if $| i|>n$.  
Let $\cn_-=\sum_{i<0} \cg_i$ (the subalgebra of strictly lower triangular matrices in $sl(n,\C)$), $\calB_+=\sum_{i\geq 0} \cg_i$, and $N_-$ and $B_+$ the  subgroups of  $SL(n,\C)$ with Lie algebra $\cn_-$ and $\calB_+$ respectively.  Define $\cn_+$, $\calB_-$, $N_+$ and $B_-$ similarly.  

Given $g\in SL(n,\C)$, the Bruhat decomposition states that we can factor $g$ uniquely as $\ell p v$, where $\ell\in N_-$ (strictly lower triangular), $v\in B_+$ upper triangular, and $p$ a permutation matrix.  The set of all $g$ with $p=$ identity is the {\it big cell\/} of $SL(n,\C)$, which is open and dense. 

The next Theorem  states that the eigen-matrix $U(\l)$ of $e_{n1}\l^n+b$ lies in the big cell of $SL(n)$ and gives the explicit formula for the Bruhat factorization of $U(\l)$. The $N_-$ factor of $U(\l)$ will be used to define an $n\times n$ analogue of the KdV reality condition.  Although the proof of the following Theorem was given in \cite{TerUhl00a}, we include the proof here to make the paper self-contained and to set up some notation.

\bthm \label{eh} Let $\a=e^{2\pi i/n}$, $a=\diag(1,\a, \ldots, \a^{n-1})$, $b=e_{12}+ e_{23} + \cdots + e_{n-1, n}$, and  
$U(\l)= ((\a^{j-1}\l)^{i-1})$. Then
\ben
\item
$U(\l)^{-1}(e_{n1}\l^n+b) U(\l) = a\l$,  
\item $U(\l)$ lies in the big cell of $SL(n)$, i.e., $U(\l)=\phi_n(\l)\psi_n(\l)$ with $\phi_n(\l)\in N_-$ and $\psi_n(\l)\in B_+$, 
\item 
$\phi_n(\l)=\sum_{i=0}^{n-1}c_i \L^i \l^i$, where
\beq\label{hf}\
\L= \sum_{k=1}^{n-1} \sigma_k e_{k+1, k}, \quad \sigma_k=\frac{1-\a^k}{1-\a}, \quad c_k= \frac{1}{\sigma_1\cdots \sigma_k}.
\eeq
\een
\ethm

\begin{proof}
It is easy to check that $U(\l)$ is the eigen-matrix for $e_{n1}\l^n+ b$, i.e., (1) holds. 

 We use (1) to get the explicit formula for $\phi_n$.   Conjugate both sides of (1) by  $\psi_n$ to get
\begin{equation}\label{bd}
\phi_n(\l)^{-1} (e_{n1}\l^n +b) \phi_n(\l) = \psi_n(\l) a\l \psi_n(\l)^{-1}. 
\end{equation}
So  the LHS of \eqref{bd} lies in $\oplus_{i\leq 1} \cg_i$ and the RHS of \eqref{bd} lies in $\oplus_{i\geq 0}\cg_i$.  Thus \eqref{bd} lies in $\cg_0+\cg_1$.  The
$\cg_0$ component of the RHS of \eqref{bd} is $a\l$ and the $\cg_1$ component of the LHS of \eqref{bd} is $b$.  Hence we have 
\begin{equation}\label{aa}
\phi_n(\l)^{-1}(e_{n1}\l^n+ b)\phi_n(\l)= a\l + b.
\end{equation} 
We get an explicit formula for $\phi_n$ from \eqref{aa}:
Let $\ad(b)$ be the Lie algebra homomorphism defined by $\ad(b)(X)=[b,X]$.   Write $\phi_n(\l)= \sum_{j=0}^{n-1} f_j \l^j$ with $f_0=\I$.  We want to solve $f_j$ from 
$$(e_{n1}\l^n + b) \phi_n(\l)= \phi_n(\l) (a\l + b).$$
Compare coefficients of $\l^j$ to get
$$\bca bf_0=f_0b,\\ bf_j = f_j b + f_{j-1}a, & {\rm if\ } 1\leq j\leq n-1.\eca$$
By assumption $f_0=\I$, so the first equation holds.  The second equation, $[b, f_1]=a$, implies that $f_1=\L$, where $\L=\sum_{k=1}^{n-1} \sigma_k e_{k+1,k}$ and $\sigma_k$ is defined by \eqref{hf}.   
A direct computation gives 
$$a\L=\a \L a, \quad a\L^k= \a^k \L^k a.$$
By induction, we get
\begin{equation}\label{ap}
[b, \L^k]= \sigma_k \L^{k-1} a.
\end{equation}
Since  $\ad(b)$ is injective on $\calB_-$, $f_k= c_k \L^k$. 
\end{proof}

\bcor
Let  $\phi_n$ be as in Theorem \ref{eh}.    Then 
\begin{equation}\label{cb}
\phi_n(\l) = C_n(\l) \phi_n(1) C_n(\l)^{-1},
\end{equation}
where   $C_n(\l)=\diag(1, \l,\,  \ldots, \l^{n-1})$.
\ecor

\begin{proof}
A direct computation shows that $C_n(\l)\L^k= \l^k \L^k C_n(\l)$.
\end{proof}

Because $\L^n=0$, we have the following proposition.
 
\bprop\label{bu} Let $\phi_n$ and $\L$ be as in Theorem \ref{eh}. Then
$\phi_n(\l)^{-1}=\sum_{j=0}^{n-1} d_j \L^j \l^j$ for some constants $d_j$ with $0\leq j\leq n-1$.
\eprop

\bdefn {\bf The $n\times n$ KdV reality condition\/}\hfil

\ss
\noindent An element  $f$ in $L(SL(n))$ or in $ \calL(sl(n))$ is said to satisfy the {\it $n\times n$ KdV reality condition\/} if 
$\phi_n(\l) f(\l) \phi_n(\l)^{-1}$ is a power series in $\l^n$, i.e.,
\begin{equation}\label{as}
\phi_n(\a\l) f(\a\l) \phi_n^{-1}(\a\l)= \phi_n(\l) f(\l) \phi_n(\l)^{-1}, 
\end{equation}  
where $\a= e^{2\pi i/n}$ and $\phi_n$ is as in Theorem \ref{eh}.
\edefn
Or equivalently, $f$ satisfies the $n\times n$ KdV reality condition if and only if $\phi_n(\l)f(\l) \phi_n(\l)^{-1}$ is a power series in $\l^n$.  We can also write this reality condition as the condition for the fixed points of an order $n$ automorphism on $L(sl(n))$:
Let $\sigma$ denote the order $n$ automorphism of $L(SL(n))$ (or $\calL(sl(n))$) defined by
$$\sigma(f)(\l)= \phi_n(\l)^{-1} \phi_n(\a\l) f(\a\l) \phi_n(\a\l)^{-1} \phi_n(\l).$$ 
  Then $f$ satisfies the $n\times n$ KdV reality condition if and only if $f$ is a fixed point of  $\sigma$. 

\bthm  \label{an}  {\bf (Splitting Theorem)}\par
Let $L(SL(n)), L_\pm(SL(n))$ be as in Example \ref{ab}, $L$ the subgroup of $f\in L(SL(n))$ satisfying the $n\times n$ KdV reality condition \eqref{as}, $L_\pm= L_\pm(SL(n))\cap L$, and  $\calL$, $\calL_\pm$ the corresponding Lie subalgebras. Then $(\calL_+, \calL_-)$ is a splitting of $\calL$.
\ethm

\begin{proof}
 It is clear that $\calL_\pm$ are subalgebras of $\calL$, $\calL_+\cap \calL_-=\{0\}$, and $\calL_++ \calL_-\subset \calL$.    For $A=\sum_i A_i \l^i$ in $\calL(sl(n))$, we use the following notations:
 $$A_+= \sum_{i\geq 0} A_i \l^i, \qquad A_-= \sum_{i<0} A_i \l^i.$$
 To prove $\calL_++\calL_-= \calL$, it suffices to prove that if $A$ is of the form
$$A(\l)=\phi_n(\l)^{-1}(\sum_{j\leq n_0}F_j\l^{nj}) \phi_n(\l),$$
 then $A_\pm\in \calL$, i.e., the $A_\pm$ satisfy the $n\times n$ KdV reality condition \eqref{as}.
 By Proposition \ref{bu},
  \begin{align*}
&\phi_n(\l)^{-1}\left(\sum_{j\geq 0} F_j \l^{nj}\right) \phi_n(\l)\in \calL_+\subset \calL, 
\\ 
&\phi_n(\l)^{-1}\left(\sum_{j\leq -2} F_j \l^{nj}\right) \phi_n(\l)\in \calL_-\subset \calL.
\end{align*}
 To prove $\calL_++\calL_-=\calL$, it is enough to prove that $(\phi_n^{-1}F_{-1}\l^{-n} \phi_n)_\pm$ are in $\calL$, i.e., they satisfies the $n\times n$ KdV reality condition.  To prove this we need the following Lemmas.

\blem Let $\L=\sum_{k=1}^{n-1} \sigma_k e_{k+1, k}$ be as in \eqref{hf}.  Then
$\L^n=0$  and $\L^j\not=0$ for all $1\leq j\leq n-1$, i.e.,  the minimal polynomial of $\L$ is $x^n=0$.
\elem

\blem\label{ar} Let $\L$ be as in \eqref{hf}. Then
the following statements are true for $1\leq k\leq n-1$:
\ben
\item $\L^k=\frac{1}{(1-\a)^k} \sum_{j=k+1}^n (1-\a^{j-1})(1-\a^{j-2})\cdots (1-\a^{j-k}) e_{j, j-k}$.
\item Let $\Tr_{k}((x_{ij})): = \sum_{j-i= k} x_{ij}$.  Then $\Tr_{-k} (\L^k)=\Tr(b^k \L^k)= \frac{n}{(1-\a)^k}$. 
\een
\elem

\begin{proof}
Since $\L= \sum_{i=1}^{j-1} \frac{1-\a^k}{1-\a} e_{k+1, k}$, (1) follows.  

To prove (2), we first claim that 
$$\L^{n-k}= b_k(c_{k-1, 0} e_{n-k+1, 1} + c_{k-1,1} e_{n-k+2, 2} + \cdots + c_{k-1, k-1} e_{n, k}),
$$
where $c_{k-1, i}$ is the coefficient of $\l^{k-1-i}$ of 
$$(\l-\a^{-1})(\l-\a^{-2}) \cdots (\l-\a^{-(k-1)})$$  
and $$b_k=\frac{n}{(1-\a)^{n-1}\sigma_{n-1} \sigma_{n-2}\cdots \sigma_{n-k+1}}, \quad \sigma_i= \frac{1-\a^i}{1-\a}.$$
We prove this claim by induction on $k$. 
Since $$1+ x+ \cdots + x^{n-1}= (x-\a)(x-\a^2)\cdots (x-\a^{n-1}),$$ set $x=1$ to get $\prod_{i=1}^{n-1} (1-\a^i)= n$.  But 
$$\L^{n-1}= \frac{(1-\a)(1-\a^2)\cdots (1-\a^{n-1})}{(1-\a)^{n-1}}\, e_{n1}= \frac{n}{(1-\a)^{n-1}}\, e_{n1}.$$ So the claim is true for $k=1$.  Suppose the claim is true for $k<n-1$.  By \eqref{ap}, $[b,\L^{n-k}]a^{-1}= \sigma_{n-k} \L^{n-(k+1)}$.   So the $(n-k+i, i+1)$-th entry of $\L^{n-k-1}$ is 
$$\frac{b_k}{\sigma_{n-k}}\ \sum_{i=0}^{k-1}(c_{k-1, i}- c_{k-1, i-1}) \a^{-i} e_{n-k+i, i+1}.$$
But
\begin{align*}
& (c_{k-1, i} -c_{k-1, i-1}) \a^{-i} \\
&=\a^{-i} \sum_{1\leq j_1< \cdots < j_i\leq k-1}(-1)^i \a^{-(j_1+ \cdots + j_i)}\\
& \quad -\a^{-i} \sum_{1\leq j_1< \cdots <j_{i-1}\leq k-1} (-1)^{i-1}\a^{-(j_1 + \cdots + j_{i-1})}\\
&= (-1)^i (\sum_{2\leq j_1< \cdots < j_i\leq k} \a^{-(j_1 + \cdots + j_{i-1})} + \sum_{2\leq j_1 <\cdots < j_{k-1}\leq k} \a^{-(j_1+\cdots + j_{i-1})}\a^{-1})\\
& = c_{k, i}.
\end{align*}
This also shows that $b_{k+1}= \frac{b_k}{\sigma_{n-k}}$.  Thus $b_k= b_1/(\sigma_{n-1}\cdots \sigma_{n-k+1})$ and  
\begin{align*}
\Tr_{n-k}(\L^{n-k})&= b_k (1-\a^{-1})\cdots (1-\a^{-(k-1)}) \\
& = \frac{n(1-\a^{-1})\cdots (1-\a^{-(k-1)})}{(1-\a)^{n-1} \sigma_{n-1}\cdots \sigma_{n-k+1}}
= \frac{n}{(1-\a)^{n-k}}.
\end{align*}
\end{proof}

\blem\label{cy} Let $\L,  b$ and $ \phi_n$ be as in Theorem \ref{eh}, $\calN_+, \calN_-$ the subalgebras of strictly upper and lower triangular matrices in $sl(n)$, and $\calB_-$ the subalgebra of lower triangular matrices in $sl(n)$.  Then we have the following.
\ben 
\item $\{\L^i b^{n-1} \L^j\n i\geq 0, j\geq 0, i+j < n-1\}$ is a basis for $\cn_+$.
\item Let $B:\calN_+\to\calN_-$ be the linear map defined by 
$$B(\L^i b^{n-1}\L^j)= \L^i b^t \L^j, \qquad i, j\geq 0, i+j< n-1.$$  Then 
\begin{align*}
\pi_+(\phi_n^{-1}F\l^{-n}\phi_n)&= -\phi_n^{-1} B(\pi_{\calN_+}(F)) \phi_n,\\
\pi_-(\phi_n^{-1} F\l^{-n} \phi_n)&= \phi_n^{-1} (B(\pi_{\calN_+}(F)) + F\l^{-n}) \phi_n,
\end{align*}
for all $F\in sl(n)$,
where $\pi_\pm$ is the projection of $\calL$ onto $\calL_\pm$ and $\pi_{\calN_+}$ is the projection onto $\calN_+$ with respect to $sl(n)=\calN_+ +\calB_-$. 
\een
\elem

\begin{proof}

By Lemma \ref{ar} and $b^{n-1}= e_{1n}$, we have   
$$\L^i b^{n-1} \L^j= d_{ij} e_{i+1, n-j}$$ for some $d_{ij}\not=0$, so (1) follows.

Recall that $\phi_n(\l)= \sum_{j=0}^{n-1} c_j \L^j\l^j$ and $\phi_n(\l)^{-1}=\sum_{j=0}^{n-1} d_j \L^j \l^j$.  So
coefficient of $\l^k$ of $\phi_n^{-1}F\l^{-n} \phi_n$ with $k\geq 0$ is 
$$\sum_{i+j=n+k} c_jd_i \L^i F\L^j.$$
If $F\in \calB_-$ and $i+j\geq n$, then the weight of $\L^i F\L^j$ is less than $-n$, which implies that $\L^iF\L^j=0$. Therefore $\pi_+(\phi_n^{-1}F\l^{-n} \phi_n)=0$ if $F\in \calB_-$.  

Use induction to prove that
\beq \label{dm}
(b+ e_{n1} \l^n)^k= (b^{n-k})^t \l^n + b^k.
\eeq
 Since $\phi_n$ and $\phi_n^{-1}$ commute with $\L^i$ and
$\phi_n$ satisfies \eqref{aa}, 
\beq\label{dl}
\phi_n(\l)^{-1}\L^i (b+ e_{n1}\l^n)^{n-1}\L^j \phi_n(\l)= \L^i (a\l + b)^{n-1}\L^j.
\eeq
But $(a\l+b)^n =\l^n$ implies that
\ben
\item[(i)] $(a\l+b)^{-1}= \l^{-n} (a\l+ b)^{n-1}$, 
\item[(ii)] $\phi_n (a\l+b)^{-1}\phi_n^{-1}= \l^{-n} (e_{n1}\l^n +b)^{n-1}$, so $(a\l+b)^{-1}$ satisfies the $n\times n$ KdV reality condition, i.e., it is in $\calL_-$,
\item[(iii)] It follows from \eqref{dl}, \eqref{dm}, and (i) that 
$$ \phi_n^{-1}\L^i (b^{n-1}+ b^t \l^n)\l^{-n} \L^j\phi_n = \L^i (a\l + b)^{-1}\L^j\, \, \in \, \calL_-$$
\een
 Given $F\in \cn_+$, since $\{\L^ib^{n-1} \L^j\n i+j<n-1\}$ is a basis for $\calN_+$, there exist $f_{ij}$ such that 
$$F= \sum_{i,j\geq 0, i+j< n-1} f_{ij} \L^i b^{n-1} \L^j.$$
(iii) implies that $\phi_n^{-1}(F\l^{-n}+B(F))\phi_n\in \calL_-$. This proves the Lemma.  
\end{proof}

This finishes the proof of Theorem \ref{an}.
\end{proof}

\ss

Let $a, b$ be in as Theorem \ref{eh}, $J= a\l +b$, and $J_j= J^j$. Then $J^n=\l^n\I_n$ and
$$\{J_j\n j>0, j\not\equiv 0 \,\, ({\rm mod\,} n)\}$$
 is a vacuum sequence for the splitting $(\calL_+, \calL_-)$ of $\calL$ given in Theorem \ref{an}.  We call the hierarchy constructed from this splitting and vacuum sequence {\it the $n\times n$ KdV hierarchy\/}. 

\ss
\ni {\bf Phase space\/}

\ss
A direct computation implies that $\calM$ defined by \eqref{by} for the $n\times n$ KdV hierarchy is equal to 
$$\calM=\{a\l + b + u\n u\in sl(n,\C)\} \cap \calL.$$  The following Proposition identifies the phase space.

\bprop\label{ad}
Let $\L, a, b$ be as in Theorem \ref{eh}, $J_1=a\l+ b$, and $\calL$ as in Theorem \ref{an}.  Then:
\ben
\item The centralizer of $\L$ in $sl(n,\C)$, $\{C\in sl(n,\C)\n C\L=\L C\}$, is equal to $\C[\L]= \sum_{i=0}^{n-1}\C \L^i$.  
\item Given $u\in sl(n,\C)$ a constant map, we have $J_1+ u\in \calL$ if and only if $u\in \C[\L]$, i.e. $u=\sum_{j=1}^{n-1}u_j \L^j$ for some $u_j\in \C$.
\een
\eprop

\begin{proof} (1) follows from a direct computation. It remains to prove (2). 
By assumption, $\phi_n(\l) u\phi_n(\l)^{-1}$ is a polynomial in $\l^n$. But it has degree at most $2n-2$ in $\l$. So $\phi_n(\l) u\phi_n(\l)^{-1}= u + C\l^n$, (i.e., $\phi_n(\l) u= (u+C\l^n) \phi_n(\l)$).  Compare the coefficients of $\l^n$ and $\l$ to conclude that $C=0$ and  $\L u= u\L$.  By (1), $u\in \C[\L]$.  But $\tr(u)=0$ and $\L^n=0$, so $u=\sum_{i=1}^{n-1} u_i \L^i$ for some $u_i$'s. 
\end{proof}

So the equations in the $n\times n$ KdV hierarchy  are partial differential equations for $\C[\L]$-valued map in $t$ and $x$ variables. 

\ms
\ni {\bf Construction of flows}

\ss

 To compute the flows in the $n\times n$ KdV hierarchy, we need to find $Q_j$'s satisfying \eqref{ag}.

\bprop\label{ca} 
Suppose $u:\R\to \C[\L]$, $a, b$ as in Theorem \ref{eh},  $\calL=\calL_+ +\calL_-$ is the splitting for the $n\times n$ KdV hierarchy, and
 $\pi_\pm:\calL\to \calL_\pm$ denote the projections of $\calL$ onto $\calL_+$ and $\calL_-$ with respect to $\calL=\calL_++\calL_-$ respectively.  Set $J=a\l +b$ and $\xi= a\l + b+ u$ with $u:\R\to \C[\L]$. 
  Then we have the following.
\ben
\item There exists unique $Q_j(\xi)$ satisfying \eqref{ag} for each $j$.
\item Write $Q_j= \sum_{i\leq j} Q_{j,i} \l^i$.  Then the flow generated by $J^j$ in the $n\times n$ KdV hierarchy is the following flow equation on $C^\infty(\R, \C[\L])$:
\begin{equation}\label{ax} 
u_{t_j}=(Q_{j, 0})_x-[b+u, Q_{j, 0}] = [a, Q_{j, -1}],
\end{equation}
This is a partial differential equation in $t_j$ and $x$. 
\item $Q_{1,0}= b+u$, $Q_{1,i}$ is a polynomial differential operator for $u$,  and $Q_{j, i}$'s are polynomials in $Q_{1, k}$'s,
\item $Q^{nk+j}= Q^j \l^{nk}$, $Q_{nk+j, i+nk}= Q_{j, i}$ for $0\leq j\leq n-1$ and $i\leq j$.  
\een

\eprop

\begin{proof}
Suppose $Q$ satisfies \eqref{ag} for $j=1$.  Then $Q^j$ satisfies \eqref{ag} for any $j$. 
Compare coefficients of $\l^i$ in $[\p_x -(a\l+b+u), Q]=0$ to get the following recursive formula
\beq\label{dn}
(Q_{j, i})_x - [b+u, Q_{j,i}]=[a, Q_{j, i-1}].
\eeq
Let 
$$J=J_1= a\l + b.$$
The $j$-th flow of the splitting $(\calL_+, \calL_-)$ and vacuum sequence $\{J^j\n j\geq 1\}$ (defined in section 1) is 
\begin{equation}\label{az}
[\p_x - (J+u), \ \ \p_{t_j} -\pi_+(Q^j)]=0.
\end{equation}
The LHS of the $j$-th flow \eqref{az} is a degree $j+1$ polynomial in $\l$, so the coefficients of $\l^i$ with $0\leq  i\leq j+1$ must be zero.  
Statement (3) was proved in Theorem 2.2 of \cite{TerUhl00a}. In fact, $Q_{1,i}$ can be computed from the recursive formula \eqref{dn} and the fact that $Q(\xi)$ is conjugate to $J_1$ and $J_1= a\l+ b$, where $a$ has distinct eigenvalues.  The third equation in \eqref{ag} is true because $Q_{1, i}(\xi)$ is a polynomial in $u$ and its derivatives and the constant term of $Q_{1,i}$ is zero for all $i\geq 0$.  

Since $\p_x-\xi$ commutes with $Q(\xi)$, $\p_x-\xi$ commutes with $(Q(\xi))^j$.  Hence $Q^j$ satisfies \eqref{ag}.  Note that $Q^n(\xi)$ is conjugate to $J^n= \l^n \I_n$. So $Q^n(\xi)= \l^n \I_n$, where $\I_n$ is the identity$n\times n$ matrix.   
Therefore statement (4) follows. 
\end{proof}

\brem 
 Because $Q^n=\l^n\I_n$, we have $(Q_1)^{nk+j}= \l^{nk} Q^j$. So we can also use two indices to label the $n\times n$ KdV flows:  The $(j, k)$-th flow is 
$$[\p_x -(a\l + b+u), \,\, \p_{t_{j,k}} - \pi_+(Q^j\l^{nk})]=0,$$
for $1\leq j\leq n-1$ and $k\geq 0$ integer.  All the flows can be written in terms of coefficients of $\l^i$'s of $Q_1(\xi)$. 
\erem

\beg {\bf The flow generated by $J^2$ in the $n\times n$ KdV hierarchy}\hfil\par

The coefficients of $\l^j$ in the expansion of 
$$Q:=Q_1=a\l + S_0+ S_{-1}\l^{-1} + S_{-2}\l^{-2} + \cdots$$ can be computed from the recursive formula \eqref{dn} and the fact that $Q$ is conjugate to $J$.  In fact, we have
\begin{align*}
& u=\sum_{j=1}^{n-1} u_j \L^j, \quad S_0= b+ u,\\
& S_{-1}= P_{-1}+ T_{-1}, \quad P_{-1}= \ad(a)^{-1}(u_x), \quad T_{-1}= u_1[b, \ad(a)^{-1}(\L)],
\end{align*}
and the second flow is 
$$\p_t u= \p_x (aS_{-1}+ S_{-1}a+ S_0^2) -[b+u, aS_{-1} + S_{-1}a]$$
with Lax pair
$$[\p_x -(a\l+ b+u), \, \p_t -(a^2\l^2 + (aS_0+S_0a)\l + (aS_{-1}+ S_{-1} a + S_0^2))]=0.$$

(1) When $n=3$, we have $\L= e_{21}+ (1+\a) e_{32}$, $\L^2= (1+\a)e_{31}$, and $u=u_1\L + u_2\L^2$, where $\a= e^{2\pi i/3}$.  The second flow in the $3\times 3$ KdV hierarchy is the following coupled non-linear Schr\"odinger equations:
\beq\label{hi}
\bca 
\p_t u_1= -\frac{\sqrt{3}}{3}\, i\,  \p_x^2 u_1 + \frac{3-\sqrt{3}\, i}{3} \p_x u_2,\\
\p_t u_2 = \frac{\sqrt{3}}{3}\, i\, \p_x^2 u_2 + 2 u_1\p_x u_1.
\eca
\eeq

(2) When $n=4$, we have $\L= e_{21}+ (1+i) e_{32} + i e_{43}$, $\L^2= (1+i) e_{31} + (i-1) e_{42}$, $\L^3= (i-1) e_{41}$, and $u= u_1 \L + u_2\L^2 + u_3 \L^3$.  The second flow in the $4\times 4$ KdV hierarchy is 
\beq
\bca
\p_t u_1= -i \,\p_x^2 u_1 +(1+i) \p_x u_2,\\
\p_t u_2= (i+1)\, \p_x u_3 + 2u_1 \p_x u_1,\\
\p_t u_3= i\, \p_x^2 u_3 + 2\p_x(u_1u_2).
\eca
\eeq

(3) For general $n$, the second flow is of the form 
$$\p_t u_k = \frac{\a^k+1}{\a^k-1}\,\p_{xx} u_k + \,\, {\rm lower\, order\, terms,}$$ where $\a= e^{2\pi i/n}$.  Note that when $n$ is even, the evolution equation for $u_{n/2}$ is of first order. The evolution equation for $u_j$'s are of second order when $j\not=n/2$.
\eeg

\beg {\bf The flow generated by $J^j$ in the $n\times n$ KdV hierarchy}
 is of the form 
$$
\frac{\p u_k}{\p t_j}= \frac{\a^{kj}-1}{(\a^k-1)^j}\, \p_x^j u_k + \, {\rm lower\, order\, terms,\,\, where \,\,\/}\a=e^{2\pi i/n}.
$$
This is because the recursive formula implies that 
$$P_{-j}(u) = -\ad(a)^{-j}(\p_x^j u) + \, {\rm lower\, order\, terms\/},$$
and the constant term of $Q^j$ is
$\sum_{i=0}^{j-1} a^{j-1-i} P_{-(j-1)} a^i$.

\eeg

\ms
\ni {\bf An equivalent splitting}\hfil
\ss

The Lie algebra $\calL$ for the $n\times n$ KdV hierarchy is isomorphic to the standard loop algebra $\calL(sl(n))$.   
 So the splitting of $\calL$  constructed in Theorem \ref{an} gives a non-standard splitting of $\calL(sl(n))$.   
 
\bthm\label{de} Let $\calL=\calL_++ \calL_-$ be the splitting for the $n\times n$ KdV hierarchy, 
$\calL(z)=\{\xi(z)=\sum_j \xi_j z^j\n \xi_j\in sl(n,\C)\}$, and $B:\cn_+\to \calN_-$ the linear map defined in Lemma \ref{cy}. Let
$\Phi:\calL(z)\to \calL$ denote the isomorphism defined by 
$$\Phi(\xi)(\l)= \phi_n(\l)^{-1} \xi(\l^n) \phi_n(\l),$$
and $\ti\calL_\pm(z):= \Phi^{-1}(\calL_\pm)$.  Then
\begin{align*}
\ti\calL_+(z)&= \{\xi(z)= \sum_{j\geq 0} \xi_j z^j\,\big|\, \xi_j\in sl(n,\C)\},\\
\ti\calL_-(z)&=\{ \xi(z)=  B((\xi_{-1})_+) + \sum_{j\leq -1}  \xi_j z^j \,\big|\, \xi_j\in sl(n,\C) \},
\end{align*}
where $(\xi_{-1})_+$ is the projection of $\xi_{-1}$ onto $\cn_+$ with respect to $sl(n,\C) = \cn_++ \calB_-$.  
\ethm

\brem
This splitting $(\ti \calL_+(z), \ti \calL_-(z))$ of $L(sl(n,\C))$ given in Theorem \ref{de} is by no means obvious. It would be interesting to give generalizations of this construction for any simple Lie algebra $\calG$. Fix a Cartan subalgebra $\calA$ and a simple root system for $\calG$. Let $\calN_+$ and $\calN_-$ denote the subalgebras spanned by positive roots and negative roots respectively.  Let $a$ be the lowest root of $\calG$, $b$ the sum of simple roots, and 
$$\hat J=az + b.$$
($z$ is the loop parameter in the algebra $\calL(\calG)$ of loops in $\calG$).    One way to generalize our construction is to find linear operator $B:\calN_+\to \calN_-$ satisfying the following conditions: 
\ben
\item $\calL_-(B):=\{\xi\in \calL(\calG)\n \xi=B((\xi_{-1})_+) +\sum_{j<0} \xi_j z^j\}$ is a Lie subalgebra of $\calL(\calG)$. Here we use $\zeta_+$ to denote the projection of $\zeta\in \calG$ onto $\calN_+$ with respect to $\calG= \calN_++\calA + \calN_-$.  Note that $\calL_-(B)$ is a Lie subalgebra of $\calL(\calG)$ if and only if
\beq\label{cd}
B(([\xi, B(\eta_+)] + [B(\xi_+), \eta])_+)=[B(\xi_+), B(\eta_+)]
\eeq
 for all $\xi, \eta\in \calG$.  Let $\calL_+(\calG)=\{ \xi\in \calL(\calG)\n \xi= \sum_{j\geq 0} \xi_j \l^j\}$.  If $B$ is an operator satisfying \eqref{cd}, then $(\calL_+(\calG), \calL_-(B))$ is a splitting of $\calL(\calG)$.  
\item The dimension of $\pi_+([J_1, \calL_-(B)])$ is equal to the rank of $\calG$.  
\item The flows generated by the splitting $(\calL_+(\calG), \calL_-(B))$ and the vacuum sequence $\{\hat J^j\n j>0\}$ are partial differential equations.  
\een 
\erem

\bs

\section{Construction of the $2n\times 2n$ KdV-II hierarchy}\label{ef}

In this section, we use a different ordering of the eigenvalues of 
$$e_{2n, 1}\l^{2n} + (e_{12} + e_{23} + \cdots + e_{2n-1, 2n})$$ and a similar construction as in section \ref{ed} to obtain a new restriction of $2n\times 2n$ AKNS hierarchy that also generalizes the KdV hierarchy.

Recall that key steps in the construction of the splitting for the $n\times n$ KdV hierarchy are:
\ben 
\item choose an ordering of eigenvalues of $e_{n1}\l^n +b$,
\item find a $N_-$-valued polynomial loop $\phi(\l)$ so that \eqref{aa} holds,
\item find an operator $B$ as in Lemma \ref{cy}.
\een
It is natural to ask what happens if we choose a different ordering of the eigenvalues of $e_{n1}\l^n+b$.
We prove in this section that for $2n\times 2n$ case if we  reorder the eigenvalues of $e_{2n1}\l^{2n}+b$ by 
$$1, \a^2, \ldots, \a^{2n}, \a, \a^3, \ldots, \a^{2n-1}$$ instead of $1, \a, \ldots, \a^{2n-1}$ (with $\a=e^{\pi i/n}$) as in section \ref{ed}, the key steps (1)-(3) for the construction of the $n\times n$ KdV hierarchy still hold and we obtain a new hierarchy. 

 Set 
\begin{align*}
&\a=e^{\frac{2\pi i}{2n}}= e^{\pi i/n}, \quad A= \diag(1, \a, \ldots, \a^{n-1}),\\
& \L= \sum_{k=1}^{n-1} \sigma_k e_{k+1, k}\, \in sl(n), \quad  \sigma_k= \frac{1-\a^{2k}}{1-\a^2},\\
&\phi(\l)= \phi_n(\l)=\sum_{k=0}^{n-1} c_k \L^k\l^k, \quad c_k= (\prod_{i=1}^k \sigma_i)^{-1}
 \end{align*}
as in Theorem \ref{eh}.  Set 
 \begin{align*}
 &\b_n= e_{12} + e_{23} + \cdots + e_{n-1, n}\in sl(n), \quad
 b= \bpm \b_n& e_{n1}\\ 0& \b_n\epm= \sum_{i=1}^{2n-1} e_{i, i+1},\\
 & \psi(\l)= \bpm \phi(\l) & 0\\ \l^n \phi(\l) & A\phi(\l) A^{-1}\epm, \quad
 \quad a=\bpm A^2&0\\ 0& \a A^2\epm. 
 \end{align*}
 Note that the matrix $a= \diag(1, \a^2, \ldots, \a^{2n}, \a, \a^3, \ldots, \a^{2n-1})$ gives a different ordering of the eigenvalues of $e_{2n,1}\l^{2n} + b$. 
 
 Theorem \ref{eh} implies that 
 $$\phi(\l) (A^2\l + \b_n)\phi(\l)^{-1}= e_{n1}\l^n + \b_n.$$
 The following Proposition can be easily checked:
 
 \bprop Let $\psi(\l), a, b$ be as above. Then
 $$\psi(\l) (a\l +b) \psi(\l)^{-1}= e_{2n,1} \l^{2n} + b.$$
 \eprop
 
 This leads us to consider the Lie subalgebra $\ti\calL$ of $\xi \in \calL(sl(2n))$ such that 
 $\psi(\l) \xi(\l) \psi(\l)^{-1}$ is a power series of $\l^{2n}$.  Set
 $$\ti\calL_\pm= \ti\calL\cap \calL_\pm(sl(2n)).$$
 The following Proposition gives the precise description of the subspace of constants in $\calL(sl(2n))$ that lies $\ti\calL$, and the proof follows from a straight forward computation:
 
\bprop\label{ei}
 Let $\psi(\l), A, \L, c_k$ be defined as above, and 
$$h_0= A, \quad h_1= \sum_{j=1}^{n-1} s_j e_{j+1, j},\quad h_j = (h_1)^j,$$
where $s_1=1$ and $s_{j+1}= \frac{\a c_{j+1}}{c_j}\, s_j$ for $1\leq j\leq n-1$.  
Then given $u\in sl(2n)$, we have $\psi(\l) u\psi(\l)^{-1}= u$ if and only if 
 $u= \sum_{i=1}^{2n-1} u_i \eta_i$ for some $u_i\in \C$, where 
$$ \eta_i =  \bpm \L^i&0\\ 0& \L^i\epm, \qquad
\eta_{n+i}= \bpm 0 & 0\\ h_i&0\epm, \quad{\rm where\,} 0\leq i\leq n-1.$$
 \eprop
 
We proceed in a similar fashion as in section \ref{ed} to prove that $(\ti\calL_+, \ti\calL_-)$ is a splitting of $\ti \calL$.  The following Lemmas analogous to Lemma \ref{cy} can be proved by straight forward computations:

\blem \label{ej}
Let $\eta_i$, $0\leq i\leq 2n-1$ be as in Proposition \ref{ei}.  Then 
$$\{\eta_i e_{1, 2n} \eta_j\n 1\leq i, j\leq 2n-1, i+j<2n\}$$ form a base of $sl(2n)$.
\elem

\blem\label{ek} Let $\psi(\l), a$ and $b$ be as above. Then
\ben
\item $(a\l+b)^{-1}= \l^{-2n} (a\l+ b)^{2n-1}$,
\item $\psi(\l) (a\l+ b)^{-1} \psi(\l)^{-1}= b^t  + e_{1, 2n} \l^{-2n}$,
\item $(a\l + b)^{-1}\in \ti\calL_-$.
\een
\elem

To prove $\ti\calL= \ti \calL_++ \ti \calL_-$, it suffices to prove the following Proposition:

\bprop Given $\xi\in sl(2n)$, write 
$$\psi(\l)^{-1}\xi \l^{-2n} \psi(\l) = \zeta_++ \zeta_-$$ 
with $\zeta_\pm \in \calL_\pm(sl(n))$.  Then $\zeta_\pm \in \ti\calL_\pm$.
\eprop

\begin{proof}
By Lemma \ref{ej}, we can write $\xi= \sum a_{ij} \eta_i e_{1, 2n} \eta_j$. By Proposition \ref{ei}, $\psi(\l)$ and 
$\eta_i$ commute. So Lemma \ref{ek} (2) implies that
$$\g_{ij}:=\psi(\l)^{-1} \eta_i (b^t + e_{1, 2n} \l^{-2n}) \eta_j\psi(\l)= \eta_i (a\l+ b)^{-1}\eta_j.$$
Since $(a\l+ b)^{-1}\in \calL_-$,  $\g_{ij}\in \ti \calL_-$.  But
\begin{align*}
&\psi(\l)^{-1}\xi \l^{-2n}\psi(\l)  = \sum a_{ij}\psi(\l)^{-1} \eta_i e_{1, 2n}\l^{-2n} \eta_j\psi(\l)\\
&= \g_{ij} -\sum a_{ij} \psi(\l)^{-1} \eta_i b^t \eta_j \psi(\l).
\end{align*}
Note that the second term lies in $\ti \calL_+$. 
\end{proof} 

Set $J=a\l +b$.  Then we can construct a hierarchy of flows for the splitting $(\ti \calL_+, \ti\calL_-)$ with vacuum sequence $\{J^i\n i\not\equiv 0 ({\rm mod\,} 2n)\}$, and call this the {\it $2n\times 2n$ KdV-II hierarchy}.    This is the usual KdV-hierarchy when $n=1$. 

\beg  {\bf The second flow in the $4\times 4$ KdV-II hierarchy}\hfil\par

We have
 \begin{align*}
 &A=\diag(1, i), \quad a= \diag(1, -1, i, -i), \quad \L= e_{21}, \quad \phi(\l)= \bpm 1& 0\\ \l & 1\epm,\\
 & \eta_1 = e_{21}+ e_{43}, \quad \eta_2= e_{31} + i e_{42}, \quad \eta_3= e_{41},\quad u= \sum_{i=1}^3 u_i \eta_i.
 \end{align*}
 The second flow in the hierarchy generated by the splitting $(\ti \calL_+, \ti \calL_-)$ for $J^2=(a\l+ b)^2$ is the following system:
 $$\bca 
 \p_t u_1= (1+i) \p_x u_2,\\
 \p_t u_2 = -i \p_{xx} u_2 + (1-i) \p_x u_3,\\
 \p_t u_3 = i \p_{xx} u_3 + 2(1+i) u_1 \p_x u_2 + (1+ i) u_2 \p_x u_1.
 \eca$$
 Its Lax pair is 
 $$[\p_x -(a\l + b+ u), \, \p_t -(a^2\l^2 + (aQ_0+ Q_0a)\l + (aQ_{-1} + Q_{-1}a + Q_0^2))]=0,$$
 where 
 $$Q_0=b+u, \quad  Q_{-1}=\ad(a)^{-1}(\p_x u) +\frac{u_1}{2}\diag(-1, 1, i, -i).$$
 \eeg

\bs

\end{document}